\documentclass{JHEP3}

\usepackage{amsmath,amssymb,graphics}

\usepackage{epsfig,multicol}

\newcommand {\nn}    {\nonumber}

\title{Localization of Matters on Pure Geometrical Thick Branes}

\author{Yu-Xiao Liu\footnote{Corresponding author.}, Xin-Hui Zhang, Li-Da Zhang and Yi-Shi Duan \\
 Institute of Theoretical Physics, Lanzhou University,
 Lanzhou 730000, P. R. China\\
 E-mail: \email{liuyx@lzu.edu.cn}, \email{zhangxingh03@lzu.cn}, \email{zhangld04@lzu.cn}, \email{ysduan@lzu.edu.cn} }

\abstract{In the literatures, several types of thick smooth brane
configurations in a pure geometric Weyl integrable 5-dimensional
space time have been presented. The Weyl geometry is a
non-Riemannian modification of 5-dimensional Kaluza--Klein (KK)
theory. All these thick brane solutions preserve 4-dimensional
Poincar\'e invariance, and some of them break $Z_2$--symmetry
along the extra dimension. In this paper, we study localization of
various matter fields on these pure geometrical thick branes,
which also localize the graviton. We present the shape of the
potential of the corresponding Schr$\mathrm{\ddot{o}}$dinger
problem and obtain the lowest KK mode. It is shown that, for both
spin 0 scalars and spin 1 vectors, there exists a continuum
gapless spectrum of KK states with $m^2>0$. But only the massless
mode of scalars is found to be normalizable on the brane. However,
for the massless left or right chiral fermion localization, there
must be some kind of Yukawa coupling. For a special coupling,
there exist a series of discrete massive KK modes with $m^2 >0$.
It is also showed that for a given coupling constant only one of
the massless chiral modes is localized on the branes.}


\begin{document}

\section{Introduction}

Recently, there has been increasing interest and considerable
activity in the study of higher-dimensional space-times with large
extra dimensions \cite{ADD,rs,Lykken}. Suggestions that extra
dimensions may not be compact
\cite{rs,Lykken,RubakovPLB1983136,RubakovPLB1983139,Akama1983,VisserPLB1985,Randjbar-DaemiPLB1986}
or large \cite{ADD,AntoniadisPLB1990} can provide new insights for
the solution of some relevant problems of high--energy physics
such as the mass hierarchy problem, dark matter, non--locality and
the cosmological constant
\cite{RubakovPLB1983139,Randjbar-DaemiPLB1986,KehagiasPLB2004}. In
the framework of brane scenarios, an important ingredient is that
gravity is free to propagate in all dimensions, whereas all the
matter fields are confined to a 3--brane with no contradiction
with present time gravitational experiments
\cite{ADD,RubakovPLB1983139,VisserPLB1985,SquiresPLB1986,gog}.

In the brane world scenario, an important question is how to
realize the brane world idea, in which a key ingredient is
localization of various bulk fields on a brane by a natural
mechanism. It is well known that the massless scalar field
\cite{BajcPLB2000} and the graviton \cite{rs} are localized on
branes of different types, and that the spin 1 Abelian vector
fields can not be localized on the Randall-Sundrum(RS) model in
five dimensions but can be localized in some higher-dimensional
cases \cite{OdaPLB2000113}. For fermions, they do not have
normalizable zero modes in five and six dimensions
\cite{BajcPLB2000,OdaPLB2000113,NonLocalizedFermion,Ringeval,KoleyCQG2005,GherghettaPRL2000,Neupane}.
Meanwhile, for the brane with inclusion of scalar backgrounds
\cite{RandjbarPLB2000} and minimal gauged supergravity
\cite{Parameswaran0608074} in higher dimensions, localized chiral
fermions can be obtained under some conditions.

Recently, thick brane scenarios based on gravity coupled to
scalars have been constructed
\cite{dewolfe,gremm,Csaki,varios,ThickBrane_Dzhunushaliev}. An
interesting feature of these models is that one can obtain branes
naturally without introducing them by hand in the action of the
theory \cite{dewolfe}. Furthermore, these scalar fields do not
play the role of bulk fields but provide the ``material" from
which the thick branes are made of. By considering a
non-Riemannian modification of 5-dimensional Kaluza--Klein (KK)
theory (in a pure geometric Weyl integrable 5-dimensional space
time), the generalized models based on gravity coupled to scalars
have been studied in Refs.
\cite{ThickBrane1,ThickBrane2,ThickBrane3}. In this scenario,
spacetime structures with pure geometric thick smooth branes
separated in the extra dimension arise. The authors obtained a
single bound state which represents a stable 4D graviton and
proved that the spectrum of massive modes of KK excitations is not
discrete or quantized at all, but continuous without mass gap due
to the asymptotic behavior of the quantum mechanics potential
\cite{ThickBrane2,ThickBrane3}. This gives an very important
conclusion: the claim that Weylian structures mimic classically
quantum behavior does not constitute a generic feature of these
geometric manifolds \cite{ThickBrane2}.



The aim of the present article is to investigate localization of
various matters on the pure geometrical thick branes obtained in
Refs. \cite{ThickBrane1,ThickBrane2,ThickBrane3}. The paper is
organized as follows: In section \ref{SecModel}, we first give a
review of the thick branes arising from a pure geometric Weyl
integrable 5-dimensional space time, which is a non-Riemannian
modification of 5-dimensional KK theory. Then, in section
\ref{SecLocalize}, we study localization of various matters on the
pure geometrical thick branes in 5 dimensions. Finally, a brief
conclusion and discussion are presented.

\section{Review of thick brane worlds arising from pure geometry}
\label{SecModel}


Let us start with a non--Riemannian generalization of KK theory,
i.e., a pure geometrical Weyl action in five dimensions
\begin{equation}
\label{action} S_5^W =\int_{M_5^W}\frac{d^5x\sqrt{-g}} {16\pi G_5}
e^{\frac{3}{2}\omega} \left[R+3\tilde{\xi}(\nabla\omega)^2 +
6U(\omega)\right],
\end{equation}
where $M_5^W$ is a 5-dimensional Weyl-integrable manifold
specified by the pair $(g_{MN},\omega)$, $g_{MN}$ is a
5-dimensional metric and $\omega$ is a Weyl scalar function. In
such manifolds the Weylian Ricci tensor is given by
$R_{MN}=\Gamma_{MN,P}^P-\Gamma_{PM,N}^P
+\Gamma_{MN}^P\Gamma_{PQ}^Q-\Gamma_{MQ}^P\Gamma_{NP}^Q$, with
$\Gamma_{MN}^P = \{_{MN}^{\;\;P}\}  - \frac{1}{2} (\omega_{,M}
\delta_N^P+\omega_{,N} \delta_M^P-g_{MN}\omega^{,P})$ the affine
connections on $M_5^W$ and $\{_{MN}^{\;\;P}\}$ the Christoffel
symbols. The parameter $\tilde{\xi}$ is a coupling constant, and
$U(\omega)$ is a self-interaction potential for $\omega$, which,
in general, breaks the invariance of the action (\ref{action})
under Weyl rescaling,
\begin{eqnarray}
\label{weylrescalings}
 g_{MN}\rightarrow\Omega^{-2} g_{MN},\;\;\;\;
 \omega\rightarrow\omega+\ln\Omega^2,\;\;\;
 \tilde{\xi} \rightarrow \tilde{\xi}/(1+\partial_\omega\ln\Omega^2)^2,
\end{eqnarray}
where $\Omega^2$ is a smooth function on $M_5^W$.
$U(\omega)=\lambda \; e^{\omega}$, where $\lambda$ is a constant
parameter, is the only functional form which preserves the scale
invariance of the Weyl action (\ref{action}). When the Weyl
invariance is broken, the scalar field transforms from a
geometrical object into an observable degree of freedom which
generates the smooth thick brane configurations, namely, $\omega$
is not a bulk field playing the role of the modulus for the extra
dimension. The Weyl action is of pure geometrical nature since the
scalar field $\omega$ enters in the definition of the affine
connections of the Weyl manifold.

The ansatz for the line-element which results in a 4-dimensional
Poincar$\acute{e}$ invariance of the Weyl action (\ref{action}) is
given by
\begin{equation}
\label{linee} ds_5^2=e^{2A(y)}\eta_{\mu\nu}dx^\mu dx^\nu + dy^2,
\end{equation}
where $e^{2A(y)}$ is the warp factor, and $y$ stands for the extra
coordinate.

In search of a solution to the setup defined by (\ref{action}) and
(\ref{linee}), we shall use the conformal technique. Via a
conformal transformation, $\hat{g}_{MN}= e^{\omega}g_{MN}$, we go
from the Weyl frame to the Riemann one, $M_5^W \rightarrow M_5^R$.
The action (\ref{action}) is mapped into the following Riemannian
form
\begin{equation}
\label{confaction}
S_5^R=\int_{M_5^R}\frac{d^5x\sqrt{-\hat{g}}}{16\pi G_5} \left[
\hat{R}+3{\xi}\left(\hat{\nabla}\omega\right)^2+6
\hat{U}(\omega)\right],
\end{equation}
where $\xi=\tilde{\xi}-1$, $ \hat{U}(\omega)=e^{-\omega}
U(\omega)$. Thus, in this frame, we have a theory which describes
5--dimensional gravity coupled to a scalar field with a
self--interaction potential. After this transformation, the line
element reads
\begin{equation}
\label{conflinee} d\hat{s}_5^2=e^{2\sigma(y)}\eta_{\mu\nu}dx^\mu
dx^\nu+e^{\omega(y)}dy^2,
\end{equation}
where $2\sigma=2A+\omega$. If we introduce a new pair of variables
$X\equiv\omega'$ and $Y\equiv 2A'$, then the field equations that
are derivable from (\ref{confaction}) with the ansatz
(\ref{conflinee}) reduce to the following pair of coupled
equations
\begin{eqnarray}
\label{fielde}
 X' + 2YX + \frac{3}{2}X^2
    &=& \frac{1}{\xi}\frac{d \hat{U}}{d\omega}e^{\omega},\\
 Y'+\frac{3}{2}XY + 2Y^2
    &=& \left(4\hat{U}-\frac{1}{\xi}\frac{d \hat{U}}{d\omega} \right) e^{\omega}.
\end{eqnarray}
As pointed out in \cite{ThickBrane1}, this system of equations can
be easily solved if one uses the restriction $X=kY$, where $k$ is
an arbitrary constant parameter which is not allowed to adopt the
value $k=-1$ because the system would be incompatible. This
condition leads to a Riemannian potential of the form
$\hat{U}=\lambda e^{\frac{4k\xi}{1+k}\omega}$. It turns out that
this constrain leads to the following simple brane configurations:

\subsection*{Configuration 1: $Z_2$--symmetric thick brane}

In this case, $-\infty<y<+\infty$ (we recall that, due to orbifold
symmetry of the solution, only one half of the extra dimension,
say $0\leq y<+\infty$, is physically relevant). The expressions
for the warp factor and the scalar field read \cite{ThickBrane1}
\begin{equation}
\label{configuration1} e^{2A(y)}=[\cosh(ay)]^b, \;\;\;
\omega=kb\ln\cosh(ay).
\end{equation}
where
\begin{eqnarray}
 a=\sqrt{\frac{4+3k}{1+k}2\lambda}, \qquad
 b=\frac{2}{4+3k}, \label{ab1}
\end{eqnarray}
and
\begin{eqnarray}
\lambda>0, \;\;\;\; k<-4/3. \label{Condition1}
\end{eqnarray}
Hence, $b$ is negative and
the warp factor is concentrated near of the origin $y=0$. The
energy density of the scalar matter is \cite{ThickBrane2}
\begin{equation}
 \mu(y)=-\frac{3 a^2 b}{4\pi G_5} (e^{ay}+e^{-ay})^{b-2}
         \left[ 1 + \frac{b}{4} (e^{ay}-e^{-ay})^2 \right].
 \label{density1}
\end{equation}
This function has two negative minima and a positive maximum at
$y=0$ between them at some $y\neq 0$, and finally it vanishes
asymptotically (see Fig. \ref{fig:EnergyDensity1}).

\FIGURE{
\includegraphics[angle=0,width=0.6\textwidth]{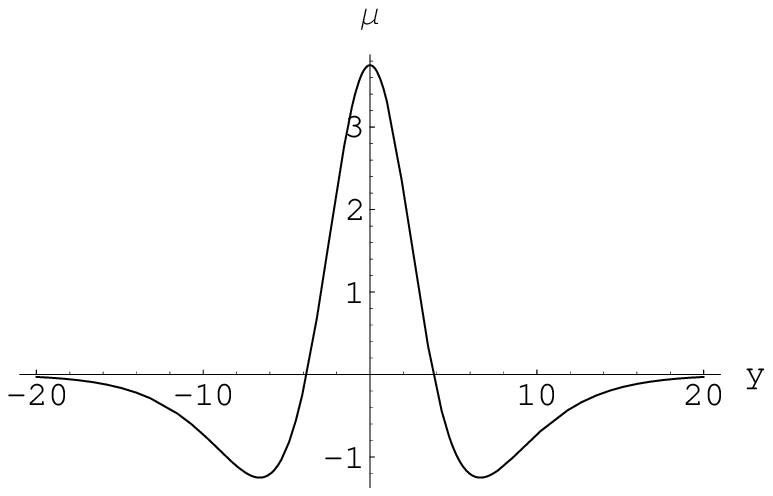}
\caption{The shape of the energy density function with $k=-5/3$
and $\lambda=0.01$. A thick brane with positive energy density is
centered at the origin $y=0$.}
\label{fig:EnergyDensity1}}

\subsection*{Configuration 2: non $Z_2$--symmetric thick brane}

The non $Z_2$--symmetric thick brane solution was found by
Barbosa-Cendejas and Herrera-Aguilar \cite{ThickBrane2}
\begin{eqnarray}
\label{configuration2}
e^{2A(y)}=k_3\left(e^{ay}+k_1e^{-ay}\right)^b,\quad
\omega=\ln\left[k_2\left(e^{ay}+k_1e^{-ay}\right)^{kb}\right],
\end{eqnarray}
where $k_2$ and $k_3$ are arbitrary constants, and
\begin{eqnarray}
\lambda>0, \;\;\;\; k<-4/3,\;\;\;\; k_1 >0. \label{Condition2}
\end{eqnarray}
The $Z_2$--symmetric solution (\ref{configuration1}) is the
particular case of this solution with $k_1=1$, $k_2=2^{-kb}$ and
$k_3=2^{-b}$. The parameter $k_1$ represents the $Z_2$--asymmetry
of the solution through a shift along the extra coordinate. This
has a quite important physical implication, i.e., the space time
is not restricted to be an orbifold geometry, it allows for a more
general type of manifolds.

\begin{figure}[htb]
\includegraphics[width=7.5cm,height=5.5cm]{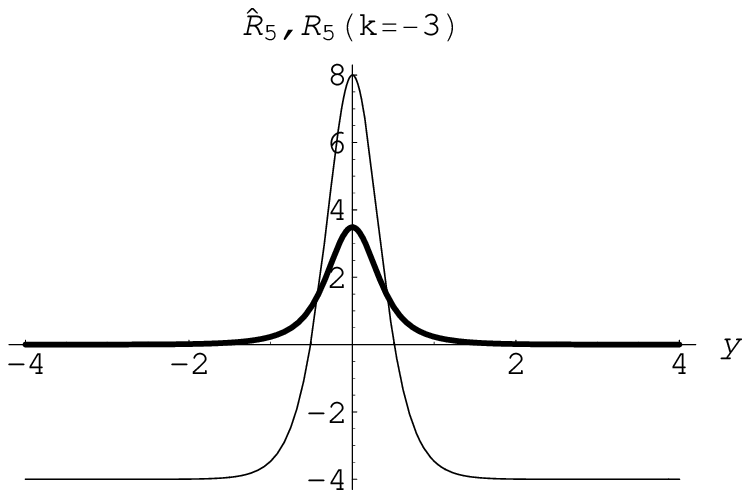}
\includegraphics[width=7.5cm,height=5.5cm]{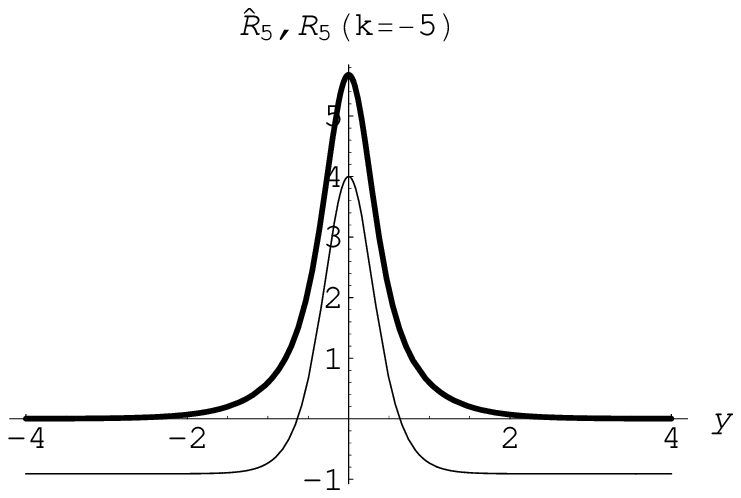}
\caption{The shape of the curvature scalar $\hat{R}_5$ in the
Riemann frame (thick line) and $R_5$ in the Weyl frame (thin
line). The parameters are set to $k_1=1$, $\lambda=1$, $k=-3$ for
left panel and $k=-5$ for right panel.}
 \label{fig_CurvatureScalar}
\end{figure}

The 5--dimensional curvature scalar in the Riemann frame and in
the Weyl frame are \cite{ThickBrane2}
\begin{eqnarray}
 \hat R_5&=&\frac{-64\lambda k_1(1+k)}{1-k}(e^{ay}+k_1e^{-ay})^{-(kb+2)} 
 \left[1+\frac{b(5+3k)}{16k_1}(e^{ay}-k_1e^{-ay})^2\right]
\end{eqnarray}
and
\begin{equation}
 R_5= \frac{-16a^2bk_1}{(e^{ay}+k_1e^{-ay})^2}
\left[1+\frac{5b}{16k_1}(e^{ay}-k_1e^{-ay})^2 \right],
\end{equation}
respectively. The shape of the curvature scalar is plotted in Fig.
\ref{fig_CurvatureScalar}. It is worth to note that both of them
are always bounded. Hence we have a 5-dimensional manifold which
is regular in both frames.

\subsection*{Configuration 3: another non $Z_2$--symmetric thick brane}

In above configurations, the parameter $\xi$ has been chosen as
$\xi=-(1+k)/(4k)$ with $k\neq -4/3$. In Ref. \cite{ThickBrane3},
the case $k=-4/3$ is consider and the corresponding solution is
read
\begin{eqnarray}
\label{configuration3}
 e^{2A}=\left[\frac{\sqrt{-8\lambda p}} {c_1}\cosh
        \left(c_1(y-c_2)\right)\right]^{\frac{3}{2p}},\quad
 \omega=-\frac{2}{p}\ln\left[\frac{\sqrt{-8\lambda
        p}}{c_1}\cosh\left(c_1(y-c_2)\right)\right],
\end{eqnarray}
where $p=1+16\xi$, $c_2$ is arbitrary integration constant, and
\begin{eqnarray}
\lambda>0, \;\;\;\; p<0,\;\;\;\; c_1>0. \label{Condition3}
\end{eqnarray}
From the solution, one can get the energy density of the scalar
matter, which behavior is similar to that of (\ref{density1}). So,
It represents a thick brane with positive energy density centered
at $y=c_2$.

These solutions would be utilized to analyze localization of
various matter fields on pure geometrical thick branes in the next
section.

\section{Localization of various matters}
\label{SecLocalize}

Now, we ask the question of whether various bulk fields with spin
ranging from 0 to 1 can be localized on thick branes by means of
only the gravitational interaction. Of course, we have implicitly
assumed that various bulk fields considered below make little
contribution to the bulk energy so that the solutions given in
previous section remain valid even in the presence of bulk fields.
The metric is given by (\ref{linee}), but it is more convenient to
change it to a conformally flat metric as
\begin{equation}
\label{conflinee2} ds_5^2=e^{2A}\left(\eta_{\mu\nu}dx^\mu
dx^\nu+dz^2\right),
\end{equation}
in which the relation of the new coordinate $z$ and $y$ is
$dz=e^{-A(y)}dy$.

\subsection{Spin 0 scalar field}

In this subsection we study localization of a real scalar field on
pure geometrical thick branes in the backgrounds
(\ref{configuration1})-(\ref{configuration3}). Let us consider the
action of a massless real scalar coupled to gravity
\begin{eqnarray}
S_0 = - \frac{1}{2} \int d^5 x  \sqrt{-g}\; g^{M N}
\partial_M \Phi \partial_N \Phi,
\label{scalarAction}
\end{eqnarray}
from which the equation of motion can be derived
\begin{eqnarray}
\frac{1}{\sqrt{-g}} \partial_M (\sqrt{-g} g^{M N} \partial_N \Phi)
= 0. \label{scalarEOM}
\end{eqnarray}
By considering the the conformally flat metric  (\ref{conflinee2})
the equation of motion (\ref{scalarEOM}) becomes
\begin{eqnarray}
\left( \partial^2_z + 3(\partial_{z} A) \partial_z
       +\eta^{\mu\nu} \partial_\mu \partial_\nu
 \right) \Phi = 0.
\end{eqnarray}

With the decomposition
\begin{eqnarray}
\Phi(x,z) =  \phi(x) \chi(z),
\end{eqnarray}
and demanding $\phi(x)$ satisfies the 4-dimensional massive
Klein--Gordon equation $(\eta^{\mu\nu}\partial_\mu \partial_\nu
-m^2 )\phi(x)=0 $, we obtain the equation for $\chi(z)$
\begin{eqnarray}
\left( \partial^2_z + 3 (\partial_{z} A) \partial_z
       +m^2 \right) \chi(z) = 0.
 \label{massiveScalar1}
\end{eqnarray}
The 5-dimensional action (\ref{scalarAction}) reduces to the
standard 4-dimensional action for the massive scalars, when
integrated over the extra dimension under the conditions that Eq.
(\ref{massiveScalar1}) is satisfied and the normalization
condition
\begin{eqnarray}
 \int^{\infty}_{-\infty} dz \;e^{3A}\chi^2(z)=1
 \label{normalizationCondition1}
\end{eqnarray}
is obeyed.

\begin{figure}[htb]
\begin{center}
\includegraphics[width=8cm,height=5.5cm]{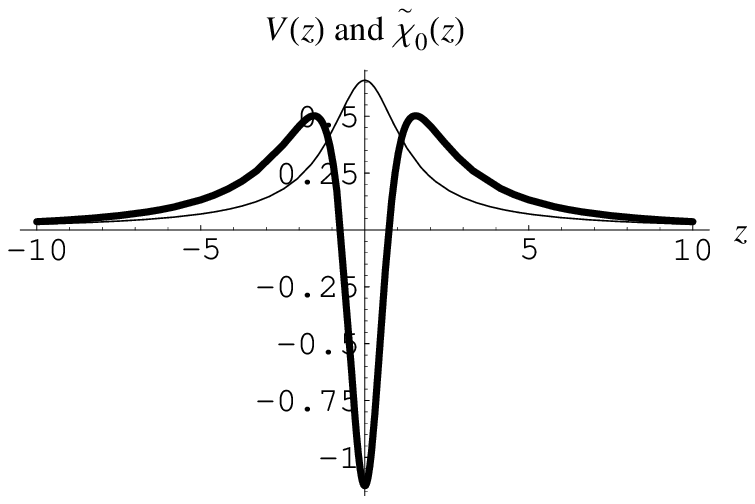}
\end{center}
\caption{The shape of the effective potential $V(z)$ (thick line)
and zero mode $\widetilde{\chi}_0(z)$ (thin line) for scalars. The
parameters are set to $k=-5/3$ and $\lambda=1/4$.}
 \label{fig_Veff_scalar}
\end{figure}

In order to obtain the Schr$\mathrm{\ddot{o}}$dinger-like
equation, we define $\widetilde{\chi}(z)=e^{\frac{3}{2}A}\chi(z)$
and get
\begin{eqnarray}
  \left[-\partial^2_z+ V(z)\right]\widetilde{\chi}(z)=m^2 \widetilde{\chi}(z),
  \label{SchEqScalar1}
\end{eqnarray}
where $m$ is the mass of the KK excitation and the potential is
given by
\begin{eqnarray}
  V(z)=\frac{3}{2}\partial^2_z A + \frac{9}{4}(\partial_z A)^2.
\end{eqnarray}
The potential depends only on the warp factor exponent $A$ and has
the same form as the case of graviton. For the first and second
brane configurations there is a particular case $k=-5/3$ for which
one can invert the coordinate transformation $dz=e^{-A(y)}dy$. For
the third brane configuration there are two particular cases
($p=-1/4$ and $p =-3/4$). In these cases we can explicitly express
$y$ in terms of $z$. Here we only consider the first brane
configuration (\ref{configuration1}), the cases of the other types
are similar. For the first case, by taking $k=-5/3$, the
expression of $y$ is $y=\mathrm{arcsinh}(az)/a$, and the effective
potential is reduced to
\begin{eqnarray}
  V(z)= \frac{9\lambda(15\lambda z^2-2)}
          {4(3\lambda z^2+1)^2}.
  \label{VeffScalar}
\end{eqnarray}
This potential has the asymptotic behavior: $V(z=\pm \infty)=0$
and $V(z=0)=-9\lambda/2$. This in fact is a volcano type potential
\cite{volcano,Davoudiasl}. This means that the potential provides
no mass gap to separate the scalar zero mode from KK modes. When
$\lambda\rightarrow \infty$, this potential tends to the singular
one found in the RS scenario \cite{rs}. The shape of the above
potential is shown in Fig. \ref{fig_Veff_scalar}. For the zero
mode $m^2=0$, the Schr$\ddot{\mathrm{o}}$dinger equation
(\ref{massiveScalar1}) can be solved. The only normalizable
eigenfunction is turned out to be
\begin{eqnarray}
  \widetilde{\chi}_0(z)=\frac{N_1}{(3\lambda z^2+1)^{3/4}},
\end{eqnarray}
where $N_1=(3\lambda)^{1/4}/\sqrt{2}$ is a normalization constant.
This function represents the lowest energy eigenfunction of the
Schr$\ddot{\mathrm{o}}$dinger equation (\ref{SchEqScalar1}) since
it has no zeros. In fact, the Schr$\ddot{\mathrm{o}}$dinger
equation (\ref{SchEqScalar1}) can be written as
$H\widetilde{\chi}=m^2\widetilde{\chi}$
\cite{dewolfe,gremm,Csaki}, where the Hamiltonian operator is
given by $H=Q^\dag Q$ with $Q=-\partial_z+(3/2)\partial_z \sigma$.
Since the operator $H$ is positive definite, there are no
normalizable modes with negative $m^2$, namely, there is no
tachyonic scalar mode. Thus the scalar zero mode is the lowest
mode in the spectrum. In addition to this massless mode, the
potential (\ref{VeffScalar}) suggest that there exists a continuum
gapless spectrum of KK modes with positive $m^2>0$, which are
similar to those obtained in Refs. \cite{Lykken,dewolfe,gremm}.

\subsection{Spin 1 vector field}

Let us turn to spin 1 vector field. Here we consider the action of
$U(1)$ vector field
\begin{eqnarray}
S_1 = - \frac{1}{4} \int d^5 x \sqrt{-g} g^{M N} g^{R S} F_{MR}
F_{NS}, \label{actionVector}
\end{eqnarray}
where $F_{MN} = \partial_M A_N - \partial_N A_M$ as usual. From
this action the equation motion is given by
\begin{eqnarray}
\frac{1}{\sqrt{-g}} \partial_M (\sqrt{-g} g^{M N} g^{R S} F_{NS})
= 0.
\end{eqnarray}
From the background geometry (\ref{conflinee2}), this equation is
reduced to
\begin{eqnarray}
 \eta^{\mu\nu} \partial_\mu F_{\nu 4} &=& 0, \\
 \partial^\mu F_{\mu \nu}
      + \left(\partial_z + \partial_{z} A \right) F_{4 \nu}&=& 0.
\label{46}
\end{eqnarray}
We assume that the $A_\mu$ are $Z_2$-even and that $A_4$ is
$Z_2$-odd with respect to the extra dimension $z$, which results
in that $A_4$ has no zero mode in the effective 4D theory.
Furthermore, in order to consistent with the gauge invariant
equation $\oint dz A_4=0$, we use gauge freedom to choose $A_4=0$.
Under these assumption, the action (\ref{actionVector}) is reduced
to
\begin{eqnarray}
S_1 = - \frac{1}{4} \int d^4 x dz
  \left(e^A \eta^{\mu\lambda} \eta^{\nu\rho}
                 F_{\mu\nu} F_{\lambda\rho}
        -2 \eta^{\mu\nu}A_{\mu}\partial_z
        \left( e^A \partial_z A_{\nu}  \right)
  \right).
\label{actionVector2}
\end{eqnarray}

By decomposing the vector field as
\begin{eqnarray}
A_{\mu}(x,z) &=& a_\mu(x)  \rho(z),
\end{eqnarray}
and importing the normalization condition
\begin{eqnarray}
 \int^{\infty}_{-\infty} dz \;e^A\rho^2(z)=1,
 \label{normalizationCondition2}
\end{eqnarray}
the action (\ref{actionVector2}) is read
\begin{eqnarray}
S_1 =  \int d^4 x
  \left( - \frac{1}{4} \eta^{\mu\lambda} \eta^{\nu\rho}
                 f_{\mu\nu} f_{\lambda\rho}
        -\frac{1}{2}m^2 \eta^{\mu\nu}a_{\mu}a_{\nu}
  \right),
\label{actionVector3}
\end{eqnarray}
where $f_{\mu\nu} = \partial_\mu a_\nu - \partial_\nu a_\mu$ is
the 4-dimensional field strength tensor, and we have required that
the $\rho(z)$ satisfy the differential equation
\begin{eqnarray}
(\partial^2_z +(\partial_{z} A) \partial_z+m^2) \rho(z) =0.
\label{diffEqVector}
\end{eqnarray}

For massive vectors, by defining $\widetilde{\rho}=e^{A/2}{\rho}$,
Eq. (\ref{diffEqVector}) changes into
\begin{eqnarray}
  \left[-\partial^2_z +V(z) \right]\widetilde{\rho}(z)=m^2
  \widetilde{\rho}(z),  \label{diffEqVector2}
\end{eqnarray}
where the potential is given by
\begin{eqnarray}
  V(z)= \frac{1}{2}\;\partial^2_z A
       +\frac{1}{4}\;(\partial_z A)^2 .
\end{eqnarray}
For the first brane configuration, by taking $k=-5/3$, the
effective potential is reduced to
\begin{eqnarray}
  V(z)=\frac{3\lambda(9\lambda z^2-2)}
          {4(3\lambda z^2+1)^2}.
  \label{VzVector}
\end{eqnarray}
The potential is very similar to the one given in Eq.
(\ref{VeffScalar}). Hence, we encounter the same analyse. The
vector zero mode is turned out to be
\begin{eqnarray}
  \widetilde{\rho}(z)=\frac{N_2}{(3\lambda z^2+1)^{1/4}},
\end{eqnarray}
where $N_2$ is a normalization constant. Now the normalization
condition (\ref{normalizationCondition2}) is read
\begin{eqnarray}
 \int^{\infty}_{-\infty} dz \;\widetilde{\rho}^{\;2}(z)
   = 1,
\end{eqnarray}
which shows that the vector zero mode is non-normalized. It is
turned out that the result is same as the RS model case, i.e. the
zero mode of the spin 1 vector field can not be localized on the
thick brane. It was shown in the RS model in $AdS_5$ space that
spin 1 vector field is not localized neither on a brane with
positive tension nor on a brane with negative tension so the
Dvali-Shifman mechanism \cite{DvaliPLB1997} must be considered for
the vector field localization \cite{BajcPLB2000}.

\subsection{Spin 1/2 fermionic field}\label{SecFermionic}

Localization of fermions in general spacetimes has been studied
for example in \cite{RandjbarPLB2000}. In Ref.
\cite{DubovskyPRD2000}, it was found that fermions can escape into
the bulk by tunneling, and the rate depends on the parameters of
the scalar field potential. In Ref. \cite{MelfoPRD2006}, Melfo
{\em et al } studied the localization of fermions on various
different scalar thick branes. They showed that only one massless
chiral mode is localized in double walls and branes interpolating
between different $AdS_5$ spacetimes whenever the wall thickness
is keep finite, while chiral fermionic modes cannot be localized
in $dS_4$ walls embedded in a $M_5$ spacetime. In this subsection
we study localization of a spin 1/2 fermionic field on the pure
geometrical thick branes.

Let us consider the Dirac action of a massless spin 1/2 fermion
coupled to gravity and scalar
\begin{eqnarray}
S_{1/2} = \int d^5 x \sqrt{-g} \left(\bar{\Psi} i \Gamma^M D_M
\Psi-\eta \bar{\Psi} F(\omega) \Psi\right), \label{DiracAction}
\end{eqnarray}
from which the equation of motion is given by
\begin{eqnarray}
\left[i\Gamma^M  ( \partial_M + \omega_M ) -\eta F(\omega)\right]
\Psi=0, \label{DiracEq1}
\end{eqnarray}
where $\omega_M= \frac{1}{4} \omega_M^{\bar{M} \bar{N}}
\Gamma_{\bar{M}} \Gamma_{\bar{N}}$ is the spin connection with
$\bar{M}, \bar{N}, \cdots$ denoting the local Lorentz indices,
$\Gamma^M$ and $\Gamma^{\bar{M}}$ are the curved gamma matrices
and the flat gamma ones, respectively, and have the relations
$\Gamma^M = e^M _{\bar{M}} \Gamma^{\bar{M}}=(e^{-A}\gamma^{\mu},-i
e^{-A}\gamma^5)$ with $e_M ^{\bar{M}}$ being the vielbein. The
spin connection $\omega_M^{\bar{M} \bar{N}}$ in the covariant
derivative $D_M \Psi = (\partial_M + \frac{1}{4} \omega_M^{\bar{M}
\bar{N}} \Gamma_{\bar{M}} \Gamma_{\bar{N}} ) \Psi$ is defined as
\begin{eqnarray}
 \omega_M ^{\bar{M} \bar{N}}
   &=& \frac{1}{2} {e}^{N \bar{M}}(\partial_M e_N ^{\bar{N}}
                      - \partial_N e_M ^{\bar{N}}) \nn \\
   &-& \frac{1}{2} {e}^{N \bar{N}}(\partial_M e_N ^{\bar{M}}
                      - \partial_N e_M ^{\bar{M}})  \nn \\
   &-& \frac{1}{2} {e}^{P \bar{M}} {e}^{Q \bar{N}} (\partial_P e_{Q
{\bar{R}}} - \partial_Q e_{P {\bar{R}}}) {e}^{\bar{R}} _M.
\end{eqnarray}
The  non-vanishing components of $\omega_M$ are
\begin{eqnarray}
  \omega_\mu = \frac{1}{2}(\partial_{z}A) \gamma_\mu \gamma_5. \label{eq4}
\end{eqnarray}
And the Dirac equation (\ref{DiracEq1}) then becomes
\begin{eqnarray}
 \left\{ i\gamma^{\mu}\partial_{\mu}
         + \gamma^5 \left(\partial_z  +2 \partial_{z} A \right)
         -\eta\; e^A F(\omega)
 \right \} \Psi =0, \label{DiracEq2}
\end{eqnarray}
where $i\gamma^{\mu} \partial_{\mu}$ is the Dirac operator on the
brane. We are now ready to study the above Dirac equation for
5-dimensional fluctuations, and write it in terms of 4-dimensional
effective fields. From the equation of motion (\ref{DiracEq2}), we
will search for the solutions of the chiral decomposition
\begin{equation}
 \Psi(x,z) = \psi_{L}(x) \alpha_{L}(z)+\psi_{R}(x) \alpha_{R}(z),
\end{equation}
where $\psi_{L}(x)$ and $\psi_{R}(x)$ are the left-handed and
right-handed components of a 4-dimensional Dirac field. Let us
assume that $\psi_{L}(x)$ and $\psi_{R}(x)$ satisfy the
4-dimensional massive Dirac equations
\begin{eqnarray}
 i\gamma^{\mu}\partial_{\mu}\psi_{L}(x)=m\psi_{R}(x),\nonumber \\
 i\gamma^{\mu}\partial_{\mu}\psi_{R}(x)=m\psi_{L}(x).\nonumber
\end{eqnarray}
Then $\alpha_{L}(z)$ and $\alpha_{R}(z)$ satisfy the following
eigenvalue equations
\begin{subequations}
\begin{eqnarray}
 \left \{ \partial_z+2\partial_{z}A
                  + \eta\;e^A F(\omega) \right \} \alpha_{L}(z)
 & = & ~~m \alpha_{R}(z), \label{CoupleEq1a}  \\
 \left \{ \partial_z+2\partial_{z}A
                  - \eta\;e^A F(\omega) \right\} \alpha_{R}(z)
 & = & -m \alpha_{L}(z). \label{CoupleEq1b}
\end{eqnarray}\label{CoupleEq1}
\end{subequations}
In order to obtain the standard four dimensional action for the
massive chiral fermions, we need the following orthonormality
conditions
\begin{eqnarray}
 \int_{-\infty}^{\infty} e^{4A}  \alpha_{L} \alpha_{R}dz
   &=& \delta_{LR}.
\end{eqnarray}
for $\alpha_{L_{n}}$ and $\alpha_{R_{n}}$.

\begin{figure}[htb]
\includegraphics[width=7.5cm,height=5.5cm]{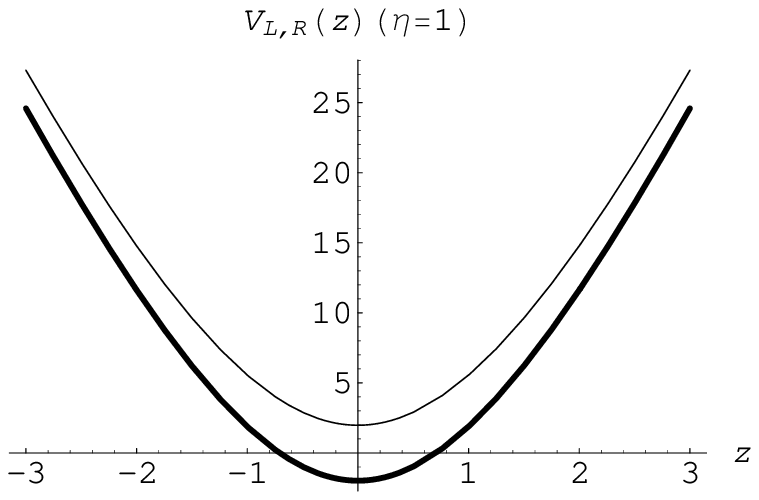}
\includegraphics[width=7.5cm,height=5.5cm]{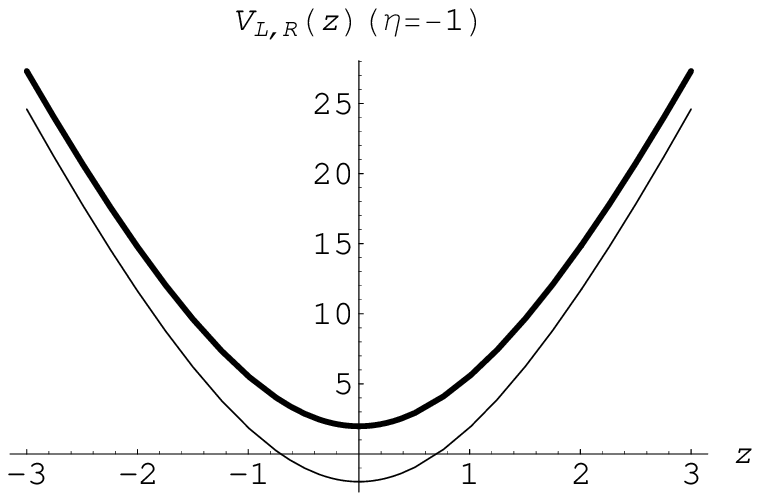}
\caption{The shape of the potentials $V_L$ and $V_R$ for left and
right chiral fermions. The parameters are set to $p=-3/4$,
$\lambda=1$, $c_1=1$, and $\eta=1$ for left panel and $\eta=-1$
for right panel. The thick line stands for the shape of the
potentials $V_L$, and the thin line for $V_R$.}
 \label{fig_Veff_Fermion}
\end{figure}

By defining $\widetilde{\alpha}_{L}=e^{2A}\alpha_{L}$, we get the
Schr$\mathrm{\ddot{o}}$dinger-like equation for the left chiral
fermions
\begin{eqnarray}
  [-\partial^2_z + V_L(z) ]\widetilde{\alpha}_{L}=m^2 \widetilde{\alpha}_{L}
  \label{SchEqLeftFermion3}
\end{eqnarray}
with the effective potential
\begin{eqnarray}
  V_L(z)&=& e^{2A} \eta^2 F^2(\omega)
     - e^{A} \eta\; \partial_z F(\omega(z))
     - (\partial_{z}A) e^{A} \eta F(\omega).
\end{eqnarray}
For localization of massive fermions around the brane, the
effective potential $V_L(z)$ should have a minimum at the brane.
Furthermore, we also demand a symmetry for $V_L(z)$ about the
position of the brane. This requires $F(\omega(z))$ to be an odd
function of $z$. So we set
$F(\omega(z))=\partial_{z}\exp{\omega(z)}$. Here we only discuss
the third configuration of brane (\ref{configuration3}) with
$p=-3/4$ and $c_2=0$  (for others configurations, the
corresponding discuss is similar). Now the potential is reduced to
\begin{eqnarray}
  V_L(z)=
   \frac{64 c_1^{14/3}\eta^2 z^2}{9 \left(c_1^4 z^2+6 \lambda \right)^{1/3}}
  -\frac{16 c_1^{7/3} \eta \left(c_1^4 z^2+9 \lambda \right)}
        {9 \left(c_1^4 z^2+6 \lambda\right)^{7/6}}.
  \label{VeffLeftFermion}
\end{eqnarray}
It is worth noting that the value of the potential at $y = 0$ is
given by
\begin{equation}
V_L(0) = -\frac{8}{3} \left(\frac{c_1}{6\lambda}\right)^{1/6}
c_1^2 \eta .
\end{equation}
For right chiral fermion, the corresponding potential is read
\begin{eqnarray}
  V_R(z)=
  \frac{64 c_1^{14/3}\eta^2 z^2}{9 \left(c_1^4 z^2+6 \lambda \right)^{1/3}}
  +\frac{16 c_1^{7/3} \eta \left(c_1^4 z^2+9 \lambda \right)}
        {9 \left(c_1^4 z^2+6 \lambda\right)^{7/6}}, \label{VeffRightFermion}
\end{eqnarray}
and the value at $y = 0$ is given by
\begin{equation}
V_R(0) = \frac{8}{3} \left(\frac{c_1}{6\lambda}\right)^{1/6} c_1^2
\eta.
\end{equation}
The shape of the above effective potentials are shown in Fig.
\ref{fig_Veff_Fermion} for different values of $\eta$. Both the
two potentials have the asymptotic behavior: $V_{L,R}(z=\pm
\infty)=\infty$. But for a given coupling constant $\eta$, the
values of the potentials at $z=0$ are opposite. For positive
$\eta$, only the potential for left chiral fermions has a negative
value at the location of the brane, which can trap the left chiral
fermion zero mode:
\begin{equation}
 \widetilde{\alpha}_{L0} \propto \exp\left\{
       -\frac{8 }{5}\eta
       \left(\frac{\sqrt{c_1^4 z^2+6\lambda}}{c_1}\right)^{5/3}
   \right\},     ~~~~~~(\eta>0)
\end{equation}
which represents the lowest energy eigenfunction of the
Schr$\ddot{\mathrm{o}}$dinger equation (\ref{SchEqLeftFermion3})
since it has no zeros. In this case, the potential for right
chiral fermions is always positive, which shows that there does
not exist zero mode. But for the case of negative $\eta$, things
are opposite and only the right chiral zero mode can be trapped on
the brane:
\begin{equation}
 \widetilde{\alpha}_{R0} \propto \exp\left\{
       \frac{8 }{5}\eta
       \left(\frac{\sqrt{c_1^4 z^2+6\lambda}}{c_1}\right)^{5/3}
   \right\}.     ~~~~~~(\eta<0)
\end{equation}
For arbitrary $\eta\neq 0$, both the two potentials suggest that
there exists a discrete spectrum of KK modes with positive
$m^2>0$, which are different form the case of the scalar obtained
in the section.

It is worth noting that, in the case of no coupling ($\eta=0$),
both the two potentials for left and right chiral fermions are
vanish, and hence there are no any localized fermion KK modes
including zero modes.

\section{Discussions}

In this paper, we have investigated the possibility of localizing
various matter fields on pure geometrical thick branes, which also
localize the graviton, from the viewpoint of field theory. We
first give a brief review of several types of thick smooth brane
configurations in a pure geometric Weyl integrable 5-dimensional
space time. Some of these thick branes break $Z_2$--symmetry along
the extra dimension.

Then, we check localization of various matter fields on these pure
geometrical thick branes from the viewpoint of field theory. For
scalars and vectors, the one dimensional
Schr$\mathrm{\ddot{o}}$dinger potentials are similar to the case
of gravity. They have a finite negative well at the location of
the brane and a finite positive barrier at each side which
vanishes asymptotically. It is shown that there is only a single
bound state (zero mode) which is just the lowest energy
eigenfunction of the Schr$\mathrm{\ddot{o}}$dinger equation for
the two kinds of fields. Since all values of $m^2>0$ are allowed,
there also exist a continuum gapless spectrum of KK states with
$m^2>0$, which turn asymptotically into continuum plane wave as
$|z|\rightarrow \infty$
\cite{Lykken,dewolfe,ThickBrane2,ThickBrane3}. But the zero mode
for spin 1 vector is non-normalized, so vector fields are not
localized on the branes. For spin $1/2$ fermion, it is shown that,
for the case of no Yukawa coupling, there is no bound states for
both left and right chiral fermions. Hence, for the massless left
or right chiral fermion localization, there must be some kind of
Yukawa coupling. These situations can be compared with the case of
the domain wall in the RS framework \cite{BajcPLB2000}, where for
localization of spin 1/2 field additional localization method by
Jackiw and Rebbi \cite{JackiwPRD1976} was introduced. In this
paper, we consider a special case of coupling as an example. With
the special coupling, we get a discrete spectrum of KK modes with
positive $m^2>0$. However, it is showed that only one massless
chiral mode is localized on the branes.

Localizing the fermionic degrees of freedom on branes or defects
requires us to introduce other interactions but gravity. Recently,
Parameswaran {\em et al} studied fluctuations about axisymmetric
warped brane solutions in 6-dimensional minimal gauged
supergravity and proved that, not only gravity, but Standard Model
fields could ``feel" the extent of large extra dimensions, and
still be described by an effective 4-Dimensional theory
\cite{Parameswaran0608074}. Moreover, there are some other
backgrounds could be considered besides gauge field
\cite{LiuJHEP2007} and supergravity \cite{Mario}, for example,
vortex background \cite{LiuNPB2007,LiuVortexFermion}. The
topological vortex coupled to fermions may result in chiral
fermion zero modes \cite{JackiwRossiNPB1981}. More recently,
Volkas {\em et al } had extensively analyzed localization
mechanisms on a domain wall. In particular, in Ref.
\cite{Volkas0705.1584}, they proposed a well-defined model for
localizing the SM, or something close to it, on a domain wall
brane. Their paper made use of preparatory work done in Refs.
\cite{GeorgePRD2007,Davies0705.1391}.

\section*{Acknowledgement}

The authors are really grateful to the referee and the authors of
Ref. \cite{ThickBrane2} for their very helpful criticisms and
recommendations which considerably improved the paper. It is also
a pleasure to thank Dr Li Zhao, Zhenhua Zhao and Shaofeng Wu for
interesting discussions. This work was supported by the National
Natural Science Foundation of the People's Republic of China (No.
502-041016, No. 10475034 and No. 10705013) and the Fundamental
Research Fund for Physics and Mathematic of Lanzhou University
(No. Lzu07002).

\end{document}